\documentclass[aps,preprint]{revtex4}%
\usepackage{amsfonts}
\usepackage{amsmath}
\usepackage{amssymb}
\usepackage{subfigure}
\usepackage{graphicx}%
\begin{document}
\preprint{CTP-SCU/2019011}
\title{Free-fall Rainbow BTZ Black Hole}
\author{Benrong Mu$^{a,b}$}
\email{benrongmu@cdutcm.edu.cn}
\author{Jun Tao$^{b}$}
\email{taojun@scu.edu.cn}
\author{Peng Wang$^{b}$}
\email{pengw@scu.edu.cn}
\affiliation{$^{a}$Physics Teaching and Research section, College of Medical Technology,
Chengdu University of Traditional Chinese Medicine ,Chengdu, 611137, PR China}
\affiliation{$^{b}$Center for Theoretical Physics, College of Physical Science and
Technology, Sichuan University, Chengdu, 610064, PR China}

\begin{abstract}
Doubly special relativity (DSR) is an effective model for encoding quantum
gravity in flat spacetime. To incorporate DSR into general relativity, one
could use \textquotedblleft Gravity's rainbow\textquotedblright, where the
spacetime background felt by a test particle would depend on its energy. For a
black hole, there are two natural orthonormal frames, the stationary one
hovering above it and freely falling one along geodesics. Since the rainbow
metric is the metric that the radiated particles \textquotedblleft
see\textquotedblright, a more natural orthonormal frame is the one anchored to
the particles. And the cases with the stationary orthonormal frame have been
extensively studied in the literature. In this paper, we investigate
properties of rainbow BTZ black holes in the scenario with the free-fall
orthonormal frame. We first review the thermodynamic properties of a BTZ black
hole. Furthermore, we obtain the Free-fall (FF) rainbow BTZ black hole and
then calculate its Hawking temperature via the Hamilton-Jacobi method.
Finally, we discuss the thermodynamic properties of a FF stationary rainbow
BTZ black hole .

\end{abstract}
\maketitle
\tableofcontents

\section{Introduction}

Shortly after Einstein proposed general relativity in 1915, Schwarzschild
derived the first black hole solution (Schwarzschild metric) from Einstein's
equations \ \cite{INT-Schwarzschild:1916uq}. From then on, the study of black
hole has become an important field of modern physics. Due to the strong
gravitational pull from a black hole, the classical theory of a black hole
predicts that anything, including light, cannot escape from the black hole.
However, considering the quantum effect, Stephen Hawking demonstrated that
black holes radiate a thermal flux of quantum particles
\cite{INT-Hawking:1974sw}.

Shortly after this discovery, it was realized that there is the
trans-Planckian problem in Hawking's calculations \cite{INT-Unruh:1976db}.
Hawking radiation appears to originate from a model with a large initial
frequency, far beyond the Planck mass $m_{p}$, which experiences an
exponentially high gravitational redshift near the horizon. Therefore,
Hawking's prediction relies on the validity of quantum field theory for
arbitrary high energy in curved spacetime. On the other hand, quantum field
theory is considered to be more of an effective field theory whose nature
remains unknown. This observation raises the question of whether any unknown
physics at the Planck scale could strongly influence Hawking radiation. It is
widely believed that trans-Planckian physics can be expressed in certain
modifications of existing models.

Although a full theory of quantum gravity has yet to available, various
theories of quantum gravity, such as loop quantum gravity, string theory, and
quantum geometry, predicts that the transformation laws of special relativity
are modified at very high energies. In particular, Deformed Special Relativity
(DSR) theory makes the Planck length a new invariant scale and guarantees the
nonlinear Lorentz transformation in the momentum spacetime
\cite{INT-AmelinoCamelia:2000ge,INT-AmelinoCamelia:2000mn,INT-Magueijo:2001cr,INT-Magueijo:2002am}%
. Specifically, the modified energy-momentum dispersion relation of a particle
of energy $E$ and momentum $p$ in DSR can take the form of
\begin{equation}
E^{2}f^{2}\left(  E/m_{p}\right)  -p^{2}g^{2}\left(  E/m_{p}\right)
=m^{2},\label{eq:MDR}%
\end{equation}
where $m_{p}$ is the Planck mass, and $f\left(  x\right)  $ and $g\left(
x\right)  $ are two general functions with the following properties:%
\begin{equation}
\lim_{x\rightarrow0}f\left(  x\right)  =1\text{ and }\lim_{x\rightarrow
0}g\left(  x\right)  =1\text{.}%
\end{equation}
The modified dispersion relation (MDR) might play an important role in
astronomical and cosmological observations, such as the threshold anomalies of
ultra high energy cosmic rays and TeV photons
\cite{INT-AmelinoCamelia:1997gz,IN-Colladay:1998fq,IN-Coleman:1998ti,IN-AmelinoCamelia:2000zs,IN-Jacobson:2001tu,IN-Jacobson:2003bn}%
. In phenomenological physics, ground observations and astrophysical
experiments have tested the predictions of MDR theory
\cite{INT-Mittleman:1999it,INT-Cane:2003wp,INT-Shao:2011uc,INT-Petry:1999fm}.
One of the most popular choice for the functions $f(x)$ and $g(x)$ has been
proposed by Amelino-Camelia et
al.\cite{IN-AmelinoCamelia:1996pj,IN-AmelinoCamelia:2008qg}, which gives
\begin{equation}
f\left(  x\right)  =1\text{ and }g\left(  x\right)  =\sqrt{1-\eta x^{n}%
}.\label{eq:AC-Dispersion}%
\end{equation}
As shown in \cite{IN-AmelinoCamelia:2008qg}, this formula is compatible with
some of the results obtained in the Loop-Quantum-Gravity approach and reflects
the results obtained in $\kappa$-Minkowski and other noncommutative
spacetimes. Phenomenological implications of this \textquotedblleft
Amelino-Camelia (AC) dispersion relation" are also reviewed in
\cite{IN-AmelinoCamelia:2008qg}.

To incorporate DSR into the framework of general relativity, Magueijo and
Smolin \cite{IN-Magueijo:2002xx} proposed the \textquotedblleft Gravity's
rainbow\textquotedblright, where the spacetime background felt by a test
particle would depend on its energy. Consequently, the energy of the test
particle deforms the background geometry and hence the dispersion relation. As
regards the metric, it would be replaced by a one parameter family of metrics
which depends on the energy of the test particle, forming a \textquotedblleft
rainbow\ metric". Specifically, for the BTZ black hole, the corresponding
\textquotedblleft rainbow metric\textquotedblright\ solution to the rainbow
Einstein's Field equations in an stationary orthonormal frame is given in
\cite{IN-Magueijo:2002xx}:%

\begin{equation}
ds^{2}=\frac{-h\left(  r\right)  }{g^{2}\left(  E/m_{p}\right)  }dt^{2}%
+\frac{\frac{dr^{2}}{h\left(  r\right)  }+r^{2}\left[  N_{\phi}\left(
r\right)  dt+d\phi\right]  ^{2}}{g^{2}\left(  E/m_{p}\right)  }.\label{SFM}%
\end{equation}
Later, Alsaleh specifically studied the thermodynamic properties of the
rainbow BTZ black hole in a stationary orthonormal frame
\cite{INT-Alsaleh:2017hlx}. \ Since the rainbow metric is the metric that the
radiated particles \textquotedblleft see\textquotedblright, a more natural
orthonormal frame is the one anchored to the particles. Actually, in section
\ref{Sec:RG} we will show that the rainbow BTZ black hole in the free-fall
orthonormal frame is given by%
\begin{equation}
ds^{2}=\left[  \frac{1}{g^{2}\left(  E/m_{p}\right)  }-\frac{1}{f^{2}\left(
E/m_{p}\right)  }\right]  h\left(  r\right)  dt^{2}+\frac{-h\left(  r\right)
dt^{2}+\frac{dr^{2}}{h\left(  r\right)  }+r^{2}\left[  N_{\phi}\left(
r\right)  dt+d\phi\right]  ^{2}}{g^{2}\left(  E/m_{p}\right)  }.\label{FFM}%
\end{equation}
In the remainder of this article, we dub the rainbow BTZ black holes $\left(
\ref{SFM}\right)  $ and $\left(  \ref{FFM}\right)  $ as stationary frames (SF)
and free fall frames (FF) rainbow BTZ black holes, respectively. In this
paper, we aim to explore the thermodynamic properties of the FF Rainbow BTZ
black hole.

There are some methods proposed to understand Hawking radiation
\cite{INT-Hemming:2001we,INT-Medved:2002zj,INT-Vagenas:2001rm,INT-Arzano:2005rs,INT-Wu:2006pz,INT-Li:2017pyf}%
. Recently, a semi-classical method of modeling Hawking radiation as a
tunneling process has been developed and attracted much attention. This method
was first proposed by Kraus and Wilczek
\cite{INT-Kraus:1994by,INT-Kraus:1994fj}, which is known as the
null geodesic method. Later, the tunneling behaviors of particles were
investigated using the Hamilton-Jacobi method
\cite{INT-Srinivasan:1998ty,INT-Angheben:2005rm,INT-Kerner:2006vu}%
. Furthermore, taking the effects of quantum gravity into account, the
Hamilton-Jacobi equation was modified, and the modified Hawking temperature
was derived  \cite{INT-Chen:2013pra,INT-Chen:2013tha,INT-Chen:2013ssa,INT-Chen:2014xsa,INT-Chen:2014xgj,INT-Mu:2015qta}. These motivate us to use the Hamilton-Jacobi method to
study the gravity rainbow effect of Hawking radiation \cite{INT-Mu:2015qna,INT-Tao:2016baz}. The cases with the
stationary orthonormal frame have been extensively studied by many authors
\cite{INT-Ling:2006az,INT-Ling:2008sy,INT-Garattini:2012ca,INT-Awad:2013nxa,INT-Ling:2005bp,INT-Galan:2006by,INT-Li:2008gs,INT-Ali:2014xqa,INT-Gim:2014ira}%
.

The BTZ black hole is a solution of Einstein field equations in three
dimensional curved space, which describes a rotating AdS geometry \cite{INT-Banados:1992wn}. It plays
an important role in Field theory and string theory. In this paper, we will
study the quantum gravity effect on BTZ black hole in the framework of gravity
rainbow theory with the free-fall orthonormal frame. The remainder of our
paper is organized as follows. In section \ref{Sec:BTZBH}, the thermodynamic
properties of BTZ black holes are briefly reviewed. In section \ref{Sec:RG},
the metric of a FF rainbow BTZ black hole is derived, and its Hawking
temperature is obtained using the Hamilton-Jacobi method. The temperature and
entropy of a FF rainbow BTZ black hole are computed. Finally, the
thermodynamic properties of the FF static rainbow BTZ black hole will be
studied. Section \ref{Sec:Con} is devoted to our discussion and conclusion.
Throughout the paper we take geometrized units $c=8G=k_{b}=1$, where the
Planck constant $\hbar$ is square of the Planck mass $m_{p}$.

\section{BTZ Black Hole}

\label{Sec:BTZBH}

The BTZ black hole is an important solution to Einstein field equation in a
$\left(  2+1\right)  $ dimensional space with a negative cosmological
constant. The action is
\begin{equation}
S=\frac{1}{2\pi}%
%TCIMACRO{\dint }%
%BeginExpansion
{\displaystyle\int}
%EndExpansion
\sqrt{-g}\left[  R+2\Lambda\right]  , \label{action}%
\end{equation}
where $\Lambda=-l^{2}$ is cosmological constant, and $l$ is AdS radius. The
BTZ black hole solution to the action $\left(  \ref{action}\right)  $ is
\cite{INT-Banados:1992wn}
\begin{equation}
ds^{2}=-h\left(  r\right)  dt^{2}+\frac{1}{h\left(  r\right)  }dr^{2}%
+r^{2}\left[  N_{\phi}\left(  r\right)  dt+d\phi\right]  ^{2},
\label{BTZmtric}%
\end{equation}
where
\begin{align}
h\left(  r\right)   &  =-M+\frac{r^{2}}{l^{2}}+\frac{J^{2}}{4r^{2}%
},\nonumber\\
N_{\phi}\left(  r\right)   &  =-\frac{J}{2r^{2}}.
\end{align}
The parameters $M$ and $J$ can be interpreted as the mass and the angular
momentum of the BTZ black hole. The BTZ black hole has two horizons located
$r=r_{\pm}$, which are determined by $h\left(  r\right)  =0$:
\begin{equation}
r_{\pm}^{2}=\frac{Ml^{2}\pm\sqrt{M^{2}l^{4}-l^{2}J^{2}}}{2}.
\end{equation}
In terms of $r_{\pm}$, we can rewrite $h\left(  r\right)  $, $M$ and $J$ as%
\begin{equation}
h\left(  r\right)  =\frac{1}{l^{2}r^{2}}\left(  r^{2}-r_{+}^{2}\right)
\left(  r^{2}-r_{-}^{2}\right)  ,M=\frac{r_{+}^{2}+r_{-}^{2}}{l^{2}}\text{ and
}J=\frac{2r_{+}r_{-}}{l}.
\end{equation}

The surface gravity $\kappa$ of the BTZ black hole at the outer horizon
$r=r_{+}$ is%

\begin{equation}
\kappa=\frac{r_{+}^{2}-r_{-}^{2}}{l^{2}r_{+}}.
\end{equation}
So the Hawking temperature $T_{h}$ of the BTZ black hole is
\cite{BBH-Chen:2012mh,BBH-Cai:1996df,BBH-Cai:1999dz,BBH-Cai:1998ep}%
\begin{equation}
T_{h}=\frac{\hbar\kappa}{2\pi}=\frac{\hbar\left(  r_{+}^{2}-r_{-}^{2}\right)
}{2\pi l^{2}r_{+}}. \label{eq:BTZT}%
\end{equation}
The first law of thermodynamics for the BTZ black hole was obtained in
\cite{INT-Bardeen:1973gs}, which reads
\begin{equation}
dM=T_{h}dS+\Omega_{H}dJ, \label{first law}%
\end{equation}
where $S$ and $\Omega_{H}=\frac{J}{2r_{+}^{2}}=\frac{r_{-}}{lr_{+}}$ are the
entropy and the angular velocity of the BTZ black hole, respectively. The eqn.
$\left(  \ref{first law}\right)  $ can be rewritten in the form:%
\begin{equation}
dS=\frac{dM-\Omega_{H}dJ}{T_{h}}=\frac{4\pi dr_{+}}{\hbar}, \label{entropy 1}%
\end{equation}
which, by integration, leads to
\begin{equation}
S=\frac{4\pi r_{+}}{\hbar}. \label{entropy2}%
\end{equation}
Note that the BTZ black hole entropy $S$ was also computed using the Euclidean
action method
\cite{INT-Banados:1992wn,BBH-Kim:1996eg,BBH-Mann:1996ze,BBH-Cai:1996df,BBH-Cai:1998ep}%
. The heat capacity of the BTZ black hole is%

\begin{equation}
C_{J}=T_{h}\left(  \frac{\partial S}{\partial T_{h}}\right)  _{J}=\frac{4\pi
r_{+}\Delta}{2-\Delta}, \label{CJ}%
\end{equation}
where $\Delta=\sqrt{1-\frac{l^{2}J^{2}}{M^{2}l^{4}}}$. Since $0\leq\Delta
\leq1$, the BTZ black hole always has a positive heat capacity, which implies
the thermodynamic system for the BTZ black hole is stable.

\section{Free-fall Rainbow BTZ Black Hole}

\label{Sec:RG}

In this section, we obtain the Free-fall (FF) rainbow BTZ black hole and then
calculate its Hawking temperature via the Hamilton-Jacobi method. Finally, we
discuss the rainbow corrections to the entropy of the black hole.

\subsection{Free-fall Rainbow BTZ Metric}

First, we generalize the energy-independent BTZ metric $\left(  \ref{BTZmtric}%
\right)  $ to the energy-dependent BTZ rainbow metric. Generally, the
energy-independent metric is given by%
\begin{equation}
ds^{2}=g_{\mu\nu}dx^{\mu}\otimes dx^{v}, \label{eq:metric}%
\end{equation}
can be rewritten in terms of a set of energy-independent orthonormal frame
fields $e_{a}$:%
\begin{equation}
ds^{2}=\eta^{ab}e_{a}\otimes e_{b},
\end{equation}
where $a=\left(  0,i\right)  $ and $i$ is the spatial index. For the MDR
$\left(  \ref{eq:MDR}\right)  $, the energy-dependent rainbow counterpart for
the energy-independent metric $\left(  \ref{eq:metric}\right)  $ can be
obtained using equivalence principle \cite{IN-Magueijo:2002xx}, which gives%
\begin{equation}
d\tilde{s}^{2}=\tilde{g}_{\mu\nu}dx^{\mu}\otimes dx^{v}=\eta^{ab}\tilde{e}%
_{a}\otimes\tilde{e}_{b}=\left(  \frac{1}{g^{2}\left(  E/m_{p}\right)  }%
-\frac{1}{f^{2}\left(  E/m_{p}\right)  }\right)  e_{0}\otimes e_{0}%
+\frac{ds^{2}}{g^{2}\left(  E/m_{p}\right)  }, \label{eq:RBMetric}%
\end{equation}
where the energy-dependent orthonormal frame fields are%
\begin{equation}
\tilde{e}_{0}=\frac{e_{0}}{f\left(  E/m_{p}\right)  }\text{ and }\tilde{e}%
_{i}=\frac{e_{i}}{g\left(  E/m_{p}\right)  }\text{.}%
\end{equation}
Note that a tilde is used for an energy-dependent quantity.

Obviously, different choices of the orthonormal frame can give different
rainbow metrics. In fact, the form of the rainbow metric crucially depends on
the time component of the orthonormal frame, $e_{0}$. For the BTZ metric
$\left(  \ref{BTZmtric}\right)  $, $e_{0}$ in the literature
\cite{INT-Alsaleh:2017hlx,FF-1-Hendi:2015bba,FF-1-Dehghani:2018qvn} is usually
chosen to be
\begin{equation}
e_{0}=\sqrt{h\left(  r\right)  }dt, \label{eq:e0SF}%
\end{equation}
where this orthonormal frame basis is hovering above the black hole. On the
other hand, the particles radiated from the black hole travel along the
geodesics. Since the rainbow metric is the metric that the radiated particles
"see", a more natural orthonormal frame is the one anchored to the particles. \

For massive particles moving along the geodesics, $\left(  e_{0}\right)
^{\mu}$ is just the 3-velocity vector $u^{\mu}$ of the geodesics. To compute
the 3-velocity vector, we consider $p_{t}$ and $p_{\phi}$, which are conserved
along geodesics (since the metric does not depend explicitly on $t$ and $\phi
$). It leads immediately to the first integrals of the $t$- and $\phi$-
equations. These are given by%
\begin{align}
u_{t}  &  =g_{tt}u^{t}+g_{t\phi}u^{\phi}=\left[  -h\left(  r\right)
+r^{2}N_{\phi}^{2}\left(  r\right)  \right]  \dot{t}+r^{2}N_{\phi}\dot{\phi
}=-E/m\label{eq:tandphi}\\
u_{\phi}  &  =g_{\phi t}u^{t}+g_{\phi\phi}u^{\phi}=r^{2}N_{\phi}\dot{t}%
+r^{2}\dot{\phi}=L/m\nonumber
\end{align}
where $m$, $E$ and $L$ are the mass, the energy and the angular momentum of
the particles. Solving eqns. $\left(  \ref{eq:tandphi}\right)  $ for $\dot{t}$
and $\dot{\phi}$ gives%
\begin{align}
u^{t}  &  =\dot{t}=\frac{N_{\phi}\left(  r\right)  L+E}{mh\left(  r\right)
},\nonumber\\
u^{\phi}  &  =\dot{\phi}=\frac{L}{mr^{2}}-\frac{N_{\phi}\left(  r\right)
\left[  N_{\phi}\left(  r\right)  L+E\right]  }{mh\left(  r\right)  }.
\end{align}
We then use $g^{\mu\nu}u_{\mu}u_{\nu}=-1$ to find $u^{r}$. So the 3-velocity
vector of the radiated particle is%
\begin{equation}
u^{\mu}=\left(  \frac{N_{\phi}\left(  r\right)  L+E}{mh\left(  r\right)
},\sqrt{\frac{\left[  E+N_{\phi}\left(  r\right)  L\right]  ^{2}}{m^{2}%
}-h\left(  r\right)  \left(  1+\frac{L^{2}}{m^{2}r^{2}}\right)  },\frac
{L}{mr^{2}}-\frac{N_{\phi}\left(  r\right)  \left[  N_{\phi}\left(  r\right)
L+E\right]  }{mh\left(  r\right)  }\right)  . \label{3-velocity}%
\end{equation}
Therefore, the time component of the orthonormal frame anchored to the
radiated particles is then given by%
\begin{equation}
e_{0}=u_{\mu}dx^{\mu}=-\frac{E+N_{\phi}\left(  r\right)  L}{m}dt+\frac
{\sqrt{\frac{\left[  E+N_{\phi}\left(  r\right)  L\right]  ^{2}}{m^{2}%
}-h\left(  r\right)  \left(  1+\frac{L^{2}}{m^{2}r^{2}}\right)  }}{h\left(
r\right)  }dr. \label{eq:e0}%
\end{equation}
From eqn. $\left(  \ref{eq:RBMetric}\right)  $, the rainbow BTZ metric is
\begin{equation}
ds^{2}=\left[  \frac{1}{g^{2}\left(  E/m_{p}\right)  }-\frac{1}{f^{2}\left(
E/m_{p}\right)  }\right]  e_{0}\otimes e_{0}+\frac{-h\left(  r\right)
dt^{2}+\frac{dr^{2}}{h\left(  r\right)  }+r^{2}\left[  N_{\phi}\left(
r\right)  dt+d\phi\right]  ^{2}}{g^{2}\left(  E/m_{p}\right)  },
\label{eq:FFRBBTZ}%
\end{equation}
where $e_{0}$ is given by eqn. $\left(  \ref{eq:e0}\right)  $. The rainbow BTZ
metric $\left(  \ref{eq:FFRBBTZ}\right)  $ is dubbed as "Free-fall (FF)
rainbow BTZ black hole."

\subsection{Effective Hawking Temperature}

We now use the Hamilton-Jacobi method to calculate the Hawking temperature of
the FF rainbow black hole $\left(  \ref{eq:FFRBBTZ}\right)  $. In the
Hamilton-Jacobi method, one ignores the self-gravitation of emitted particles
and assumes that their action satisfies the relativistic Hamilton-Jacobi
equation. The tunneling probability for the classically forbidden trajectory
from inside to outside the horizon is obtained by using the Hamilton-Jacobi
equation to calculate the imaginary part of the action for the tunneling
process. In \cite{INT-Mu:2015qna}, it showed that, in the rainbow metric
$ds^{2}=\tilde{g}_{\mu\nu}dx^{\mu}dx^{\nu}$, the Hamilton-Jacobi equations for
massive particles can simply be written as%
\begin{equation}
\tilde{g}_{\mu\nu}\partial^{\mu}I\partial^{\nu}I=m^{2},
\label{eq:Hamilton-Jacobi}%
\end{equation}
where $I$ is the radiated particle's classical action. Since the FF rainbow
black hole $\left(  \ref{eq:FFRBBTZ}\right)  $ is independent of $t$ and
$\phi$, we can employ the following ansatz for the action $I$%
\begin{equation}
I=-Et+W\left(  r\right)  +L\phi,
\end{equation}
where, as above defined, $E$ and $L$ are the radiated particle's energy and
angular momentum, respectively. Defining $p_{r}\left(  r\right)  \equiv
W^{\prime}\left(  r\right)  $, we can use the FF rainbow black hole $\left(
\ref{eq:FFRBBTZ}\right)  $ to rewrite the Hamilton-Jacobi equation $\left(
\ref{eq:Hamilton-Jacobi}\right)  $ as the equation in terms of $p_{r}$, which
is too long to put here. Solving the Hamilton-Jacobi equation for $p_{r}$, we
obtain%
\begin{equation}
p_{r}^{\pm}\left(  r\right)  =\frac{A_{r}^{\pm}\left(  r\right)  }{h\left(
r\right)  \tilde{H}\left(  r\right)  } \label{eq:RBpr}%
\end{equation}
where $+$/$-$ denotes the outgoing/ingoing solutions. Here, $A_{r}^{\pm
}\left(  r\right)  $ are regular functions of $r$ without poles, the detailed
forms of which are rather complicated and not relevant. In the denominator of
$p_{r}^{\pm}\left(  r\right)  $, we define%
\begin{align}
\tilde{H}\left(  r\right)   &  \equiv h\left(  r\right)  -\epsilon\left(
E/m_{p}\right)  \frac{\left[  E+LN_{\phi}\left(  r\right)  \right]  ^{2}%
}{m^{2}}\text{,}\nonumber\\
\epsilon\left(  E/m_{p}\right)   &  \equiv1-\frac{g^{2}\left(  E/m_{p}\right)
}{f^{2}\left(  E/m_{p}\right)  }.
\end{align}
The corresponding action is%
\begin{equation}
I_{\pm}=-Et+\int p_{r}^{\pm}\left(  r\right)  dr+L\phi,
\end{equation}
where the imaginary part of $I_{\pm}$ comes from the integral of $p_{r}^{\pm
}\left(  r\right)  $.

According to the Hamilton-Jacobi method, the residues of $p_{r}^{\pm}\left(
r\right)  $ lead to the Hawking temperature of the radiation. So the poles of
$p_{r}^{\pm}\left(  r\right)  $ correspond to the locations of the horizons of
the FF rainbow black hole $\left(  \ref{eq:FFRBBTZ}\right)  $. Eqn. $\left(
\ref{eq:RBpr}\right)  $ shows that the poles of $p_{r}^{\pm}\left(  r\right)
$ are at the location where $h\left(  r\right)  =0$ or $\tilde{H}\left(
r\right)  =0$. For $h\left(  r\right)  =0$, one simply has $r=r_{\pm}$.
Moreover, it can show that there are also two solutions to $\tilde{H}\left(
r\right)  =0$, namely $r=\tilde{r}_{\pm}$ with $\tilde{r}_{+}\geq\tilde{r}%
_{-}$. The ordering of $r_{\pm}$ and $\tilde{r}_{\pm}$ depends on the sign of
$\epsilon\left(  E/m_{p}\right)  $. Since $h^{\prime}\left(  r\right)  >0$ for
$r\geq r_{+}$ and $h^{\prime}\left(  r\right)  <0$ for $r\leq r_{-}$, we then
have $\tilde{r}_{+}\geq r_{+}\geq r_{-}\geq\tilde{r}_{-}$ when $\epsilon
\left(  E/m_{p}\right)  >0$. In this case, $\tilde{r}_{+}$ is the radius of
the outermost horizon. Similarly, in the case with $\epsilon\left(
E/m_{p}\right)  <0$, the radius of the outermost horizon is $r_{+}$.

To find the Hawking temperature on the outermost horizon, we need to calculate
the imaginary part of $I_{\pm}$ by integrating $p_{r}^{\pm}\left(  r\right)  $
along the semicircle around the outermost horizon. As shown in
\cite{FF-2-Tao:2017mpe,FF-2-Wang:2015zpa}, the probability of a particle
tunneling from inside to outside the horizon is%
\begin{equation}
P_{emit}\propto\exp\left[  -\frac{2}{\hbar}\left(  \operatorname{Im}%
I_{+}-\operatorname{Im}I_{-}\right)  \right]  .
\end{equation}
For a particle of energy $E$ and angular momentum $L$ residing in a system
with temperature $T$ and angular velocity $\omega$, the Maxwell--Boltzmann
distribution is \cite{FF-2-landau}
\begin{equation}
P\propto\exp\left[  -\frac{E-\omega L}{T}\right]  \text{.} \label{eq:MBD}%
\end{equation}
From eqn. $\left(  \ref{eq:MBD}\right)  $, the effective Hawking temperature
can be read off from the Boltzmann factor in $P_{emit}$:
\begin{equation}
\tilde{T}_{h}=\frac{\hbar\left[  E+N_{\phi}\left(  r_{h}\right)  L\right]
}{2\left(  \operatorname{Im}I_{+}-\operatorname{Im}I_{-}\right)  },
\label{eq:Eff-Temp}%
\end{equation}
where $-N_{\phi}\left(  r_{h}\right)  =-\tilde{g}_{t\phi}/\tilde{g}_{\phi\phi
}=-g_{t\phi}/g_{\phi\phi}$ is the angular velocity of the FF rainbow BTZ black
hole, and $r_{h}=r_{+}$ or $\tilde{r}_{+}$ is the outermost horizon. For the
$\epsilon\left(  E/m_{p}\right)  <0$ and $\epsilon\left(  E/m_{p}\right)  >0$
cases, we find

\begin{itemize}
\item $\epsilon\left(  E/m_{p}\right)  <0$: The outermost horizon is at
$r=r_{+}$. Using the residue theory for the semi circle around $r=r_{+}$, we
get%
\begin{equation}
\operatorname{Im}I_{+}=\operatorname{Im}I_{-}=\frac{\pi\left[  E+N_{\phi
}\left(  r_{+}\right)  L\right]  }{h^{\prime}\left(  r_{+}\right)  },
\end{equation}
which gives the effective Hawking temperature
\begin{equation}
\tilde{T}_{h}=\infty.
\end{equation}

\item $\epsilon\left(  E/m_{p}\right)  >0$: The outermost horizon is at
$r=\tilde{r}_{+}$. Using the residue theory for the semi circle around
$r=\tilde{r}_{+}$, we get%
\begin{align}
\operatorname{Im}I_{+}  &  =0\\
\operatorname{Im}I_{-}  &  =-\frac{2\pi\left[  E+N_{\phi}\left(  r_{+}\right)
L\right]  }{\tilde{H}^{\prime}\left(  \tilde{r}_{+}\right)  }\sqrt{\frac
{g^{2}\left(  E/m_{p}\right)  \left(  1+\frac{L^{2}}{m^{2}\tilde{r}_{+}^{2}%
}\right)  }{f^{2}\left(  E/m_{p}\right)  }-\frac{L^{2}}{m^{2}\tilde{r}_{+}%
^{2}}},
\end{align}
which gives the effective Hawking temperature
\begin{equation}
\tilde{T}_{h}=\frac{\hbar\tilde{H}^{\prime}\left(  \tilde{r}_{+}\right)
}{4\pi\sqrt{1-\epsilon\left(  E/m_{p}\right)  \left(  1+\frac{L^{2}}%
{m^{2}\tilde{r}_{+}^{2}}\right)  }}. \label{eq:FFRBT}%
\end{equation}

When $g\left(  E/m_{p}\right)  =f\left(  E/m_{p}\right)  =1$, one has
$\epsilon\left(  E/m_{p}\right)  =0$, $\tilde{r}_{+}=r_{+}$ and $\tilde
{H}\left(  r\right)  =h\left(  r\right)  $, which means that, as expected,
$\tilde{T}_{h}$ of the FF rainbow BTZ black hole would reduce to $T_{h}$ of
the BTZ black hole, given in eqn. $\left(  \ref{eq:BTZT}\right)  $. To express
$\tilde{T}_{h}$ in terms of $E$ and $L$, we need to solve $\tilde{H}\left(
r\right)  =0$ for $\tilde{r}_{+}$, whose expression is quite complicated.
However for $\epsilon\left(  E/m_{p}\right)  \ll1$, one has that, to
$\mathcal{O}\left(  \epsilon\left(  E/m_{p}\right)  \right)  $,
\begin{equation}
\tilde{T}_{h}\approx T_{h}\left\{  1+\frac{\epsilon\left(  E/m_{p}\right)
}{2}\left[  1+\frac{L^{2}}{m^{2}r_{+}^{2}}+\frac{\hbar^{2}\left(  E-\frac
{JL}{2r_{+}^{2}}\right)  ^{2}\left(  \frac{2}{l^{2}}+\frac{3J^{2}}{2r_{+}^{4}%
}\right)  }{8\pi^{2}T_{h}^{2}m^{2}}-\frac{\hbar\left(  E-\frac{JL}{2r_{+}^{2}%
}\right)  LJ}{\pi T_{h}m^{2}r_{+}^{3}}\right]  \right\}  .
\end{equation}
Since the Hawking radiation spectrum is dominated by low angular momentum
modes \cite{FF-2-Caughey:2016mnq}, we can set $L=0$ for simplicity. In this
case, the effective Hawking temperature becomes%
\begin{equation}
\tilde{T}_{h}\approx T_{h}\left\{  1+\frac{\epsilon\left(  E/m_{p}\right)
}{2}\left[  1+\frac{\hbar^{2}E^{2}\left(  \frac{2}{l^{2}}+\frac{3J^{2}}%
{2r_{+}^{4}}\right)  }{8\pi^{2}T_{h}^{2}m^{2}}\right]  \right\}  ,
\end{equation}
which shows that the rainbow gravity correction tends to increase the Hawking
temperature of the black hole.
\end{itemize}

\subsection{Thermodynamics of FF Static Rainbow BTZ Black Hole}

For simplicity, we estimate the rainbow corrected temperature and entropy for
a FF static rainbow BTZ black hole, which has $J=0$. First, we can use the
Heisenberg uncertainty principle to estimate the momentum $p$ of an emitted
particle \cite{FF-3-TRSBH-Bekenstein:1973ur,FF-3-TRSBH-Adler:2001vs}:
\begin{equation}
p\sim\delta p\sim\hbar/\delta x\sim\hbar/\tilde{r}_{+}%
.\label{eq:momentumhorizon}%
\end{equation}
Using $\tilde{H}\left(  \tilde{r}_{+}\right)  =0$ and eqns. $\left(
\ref{eq:momentumhorizon}\right)  $ and $\left(  \ref{eq:MDR}\right)  $, we can
use express the energy $E$ in terms of the black hole mass $M$. The effective
Hawking temperature $\left(  \ref{eq:FFRBT}\right)  $ then can be written as a
function of $M$, $T\left(  M\right)  $, which can be intepreted as the rainbow
corrected temperature of the black hole. Using the first law of black hole
thermodynamics, we find that the entropy of the black hole is%
\begin{equation}
S\left(  M\right)  =\int\frac{dM}{T\left(  M\right)  }.
\end{equation}
When $M\gg\frac{\hbar^{2}}{m^{2}l^{2}}$, the corrected Hawking temperature and
entropy are estimated as
\begin{align}
T\left(  M\right)   &  \sim\frac{\hbar\sqrt{M}}{2\pi l}\left[  1+\frac
{\epsilon\left(  m/m_{p}\right)  \left(  M+1\right)  }{2M}\right]
,\nonumber\\
S\left(  M\right)   &  \sim\frac{\hbar M^{\frac{3}{2}}}{3l\pi}\left[
1+\frac{\epsilon\left(  m/m_{p}\right)  \left(  M+3\right)  }{2M}\right]  ,
\end{align}
respectively.

\begin{figure}[tb]
\begin{center}
\includegraphics[width=0.48\textwidth]{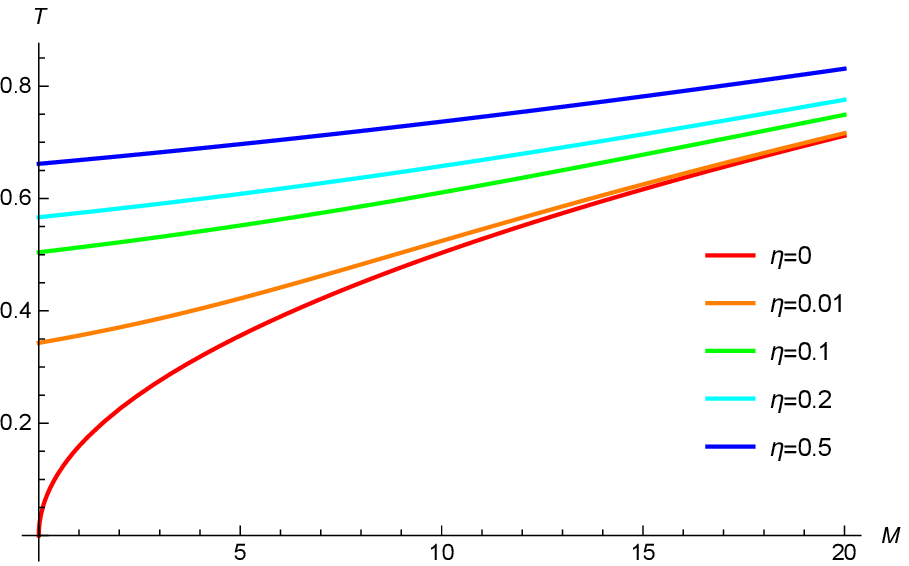}
\includegraphics[width=0.48\textwidth]{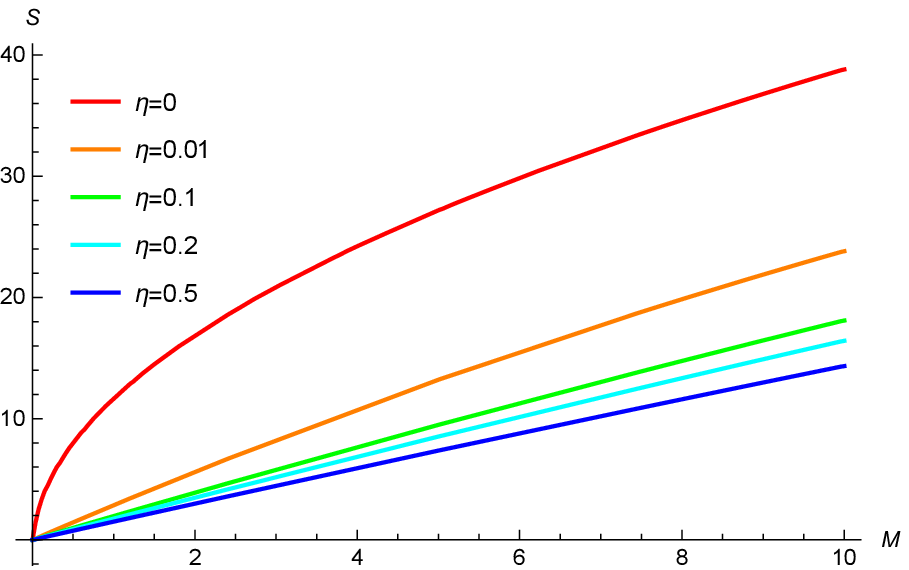}
\end{center}
\caption{{\footnotesize Plot of the corrected temperature $T\left(  M\right)
$ and entropy $S\left(  M\right)  $ for various values of $\eta.$}}%
\label{fig:TS}%
\end{figure}

To numerically investigate $T\left(  M\right)  $ and $S\left(  M\right)  $, we
focus on the Amelino-Camelia dispersion relation $\left(
\ref{eq:AC-Dispersion}\right)  $ with $n=2$ and $\eta>0$. We plot $T\left(
M\right)  $ and $S\left(  M\right)  $ for various values of $\eta$ in FIG.
\ref{fig:TS}, where we take $m=0.01$. The left panel of FIG. \ref{fig:TS}
shows that the black hole temperature increases with increasing $\eta$, which
implies that the rainbow effects would speed up the evaporation of the black
hole. Moreover, the terminal temperature is zero when $\eta=0$ while it is
greater than zero when $\eta>0$, which means the rainbow effects would lead to
a more violent death of the black hole. On the other hand, the right panel of
FIG. \ref{fig:TS} shows the black hole entropy decreases with increasing
$\eta$. Therefore, the black hole tends to store less information when the
rainbow effects are turned on.

\section{Discussion and Conclusion}

\label{Sec:Con}

In this paper, we considered a FF rainbow BTZ black hole and analyzed the
effects of rainbow gravity on the temperature and entropy. We first derived
the metric of a FF rainbow BTZ black hole. Then, we used the Hamilton-Jacobi
method to obtain the effective Hawking temperature of the rainbow BTZ black
hole, which was shown to depend on the energy of radiated particles. We found
that, when $\epsilon\left(  E/m_{p}\right)  >0^{\left[  \ref{ft:1}\right]  }%
$\footnotetext[1]{\label{ft:1} Note that, for massless particles, one has%
\[
\frac{E}{p}=\sqrt{1-\epsilon\left(  E/m_{p}\right)  }\text{,}%
\]
which means that $\epsilon\left(  E/m_{p}\right)  >0$ corresponds to the
subluminal case that has $E/p<1$, and that $\epsilon\left(  E/m_{p}\right)
<0$ to the superluminal case that has $E/p>1$.}, the radiated particles
experience a infinity Hawking temperature, which might shed light on the black
hole firewall paradox. Finally, we employed the uncertainty principle to
estimate the corrected temperature and entropy of the black hole. It showed
that, for when $\epsilon\left(  E/m_{p}\right)  <0$, the black hole evaporates
faster and stores less information than in the usual case.

\begin{acknowledgments}
We are grateful to Dr. H. Wu, and Prof. D. Chen for their useful discussions.
This work is supported in part by NSFC (Grant No.11747171, 11005016, 11175039
and 11375121) and the Fundamental Research Funds for the Central Universities.
Natural Science Foundation of Chengdu University of TCM (Grants nos. ZRYY1729
and ZRQN1656). Discipline Talent Promotion Program of /Xinglin Scholars(Grant
no. QNXZ2018050) and the key fund project for Education Department of Sichuan
(Grant no. 18ZA0173)
\end{acknowledgments}

\end{document}